\documentclass[floatfix,preprintnumbers,superscriptaddress,showpacs,
               showkeys,prl,preprint]{revtex4}

\usepackage{graphicx}
\usepackage{dcolumn}
\usepackage{amsmath, amsfonts, amssymb, bm}

\begin{document}

\bibliographystyle{apsrev} 
\preprint{CSIC/08/ICMM/graf-S-pj-05}

\title{Ab-initio calculation of the
effect of stress on the chemical activity of 
graphene} 

\author{P.L. de Andres}
\author{J.A. Verg{\'e}s}

\affiliation{
Instituto de Ciencia de Materiales de Madrid (CSIC)
E-28049 Cantoblanco, Madrid (SPAIN)
}

\date{\today}

\begin{abstract}
Graphene layers are stable, hard, and relatively inert. 
We study how tensile stress affects $\sigma$ and $\pi$ bonds and the 
resulting change in the chemical activity. 
Stress affects more strongly $\pi$ bonds that can become chemically active 
and bind to adsorbed species more strongly.
Upon stretch, single C bonds are activated in a geometry mixing 
$120^{o}$ and $90^{o}$; an intermediate state between 
$sp^{2}$ and $sp^{3}$ bonding. 
We use ab-initio density functional theory to study the adsorption 
of hydrogen on large clusters and 2D periodic models for graphene. 
The influence of the exchange-correlation functional on the adsorption 
energy is discussed.
\end{abstract}




\maketitle

The recent experimental ability to produce and characterize systems formed with
few graphene layers (FGL), down to a single graphene layer, has opened up
interesting horizons related to carbon-based materials. 
Elegant experiments on FGL have produced remarkable and unexpected 
results\cite{novoselov04,geim07}, in particular measurements of high mobilities
for carriers, raising hopes about faster electronic 
devices\cite{obradovic06,wang08}. 
Central to those applications is the ability to dope the material so
its electronic structure can be controlled to make it useful.
This doping can be obtained by different methods, e.g., as an effect from
the supporting substrate\cite{zhou07}, 
by adsorption/substitution of appropriated donors/acceptors\cite{bena08}, 
by taking advantage of external/internal stresses\cite{deandres08}, etc.
To fully realize this potential a thorough understanding of adsorption
of atoms and molecules on FGL, including all the atomic structural 
consequences, seems necessary. 
This, in turn, may be interesting for other problems, 
like the ability of graphene sheets to detect adsorbed molecules
down to very low concentrations\cite{schedin06}.
Finally, the question of whether graphene layers are flat 
or corrugated\cite{fasolino07}, at which
scale, and why, is intertwined at the atomic scale with the role of 
impurities on the layer. Even for light impurities as H, our simulations 
predict a long range modulation of the lattice of $\sim 0.05$ {\AA} on 
distances of $\sim 10$ {\AA} (the largest one we have included in our 
{\it ab-initio} simulations).

Chemistry of graphene layers can be pictured in terms of the formation of 
$\sigma$ and $\pi$ bonds. Electrons in $\sigma$ bonds occupy bonding 
combinations of $sp^{2}$ orbitals resulting from the hybridization of 
$s$, $p_x$, and $p_y$ atomic states. Each C atom contributes three electrons 
to $\sigma$ bonds that can be seen as localized along C-C directions forming
angles of 120$^{o}$ to minimize electrostatic repulsion among 
electrons. On the other hand, $\pi$ bonding results from the occupation of 
extended orbitals coming from the hybridization of $p_z$ atomic orbitals over 
the whole layer. One electron per C atom is allocated in $\pi$ orbitals giving 
rise to a total bond order of $1\frac{1}{3}$\cite{pauling}. This scenario 
predicts the formation of a stable and hard layer with a honeycomb geometry 
displaying little chemical activity due to the efficiency of this regular 
planar arrangement to maximize the bond order for the available number of 
valence electrons of carbon. Indeed, the formation of a non-planar structure 
like the one found in diamond, based in $sp^{3}$-like orbitals, where each 
carbon has four neighbours and the bond order goes down to $\sim 1$ is 
meta-stable at 0 K and 0 GPa with respect to the stacking of graphene layers 
bound together by weak van der Waals forces.
The interest of controlling the chemical activity of graphene layers cannot be 
overstated. To transform the almost inert layer into an active one we analyze 
the effect of internal/external stress on the hybridrization giving raise to 
$\pi$ orbitals. Weakening of hoppings giving rise to extended $\pi$ orbital 
results in the appearance of one electron localized on a carbon $p_z$-like 
orbital; this electron becomes available to form a single covalent bond at 
90$^{o}$ with the layer, while the $\sigma$ bonds are weakened and become 
longer, but keeping their $120^{o}$-planar arrangement. This intermediate kind 
of bond is not $sp^{3}$-like yet, but it can be considered a 
precursor since it is based on single bonds only\cite{deandres08}.
We notice that physically this picture is made possible because hoppings 
related to the formation of the $\pi$ state are 
(i) smaller than the ones related to the formation of the $sp^{2}$ orbital 
by nearly a factor of two, and 
(ii) for the relevant distances, i.e., intermediate between 
a double and a single carbon bond, they decay faster with 
distance by
$\frac{t_{pp\pi}}{t_{ss\sigma}}=6e^{-2.3 r}$
\cite{porezag95}. 
The use of mechanical forces to reshape the chemical activity is in fact
a mature field\cite{beyer05}. 
In particular, it has been reported that external stresses 
of around $0.5$ GPa can largely modify atomic bonds on carbon-based 
polymers\cite{vettegren73}. 
The stress distribution can be very inhomogeneous, 
resulting at the atomic level in a few particular bonds experiencing
local stresses $10-100$ larger than the applied external 
ones. Recently, the elastic properties of single graphene layers have
been measured showing how these layers only break for loads larger
than $42$ N m$^{-1}$ producing around 25\% elongation of the
C-C distance on the layer\cite{lee08}. 
Carbon nanotubes show similar elastic behaviour\cite{yu00}
and similar ideas should apply to these, although the existence of
a small constant curvature makes the interpretation a bit more involved
than for 2D planar graphene sheets.

We substantiate these ideas by computing total energies using 
ab-initio density functional theory (DFT)\cite{hohenberg64}. 
Both finite clusters and extended periodic systems have been 
used as models. For clusters we use localized linear combinations of atomic 
orbitals\cite{gamess} and a hybrid functional (B3LYP\cite{b3lyp})
while for periodic boundary conditions the chemistry is based on
plane-waves\cite{payne05,accelrys} and local density
approximation (LDA) or gradient corrected exchange-correlation
functionals (PBE)\cite{pbe}.
Clusters made of 50 to 100 atoms have been found adequate
regarding its size (Fig.~\ref{FigHC73H22}) 
while the infinite graphene layer has been
described using a $n \times n$ super-cell 
($n=4, 6$) and a 20 {\AA} separation to minimize 
interactions in the direction perpendicular to the 
layer (Fig.~\ref{Fig6x6}).
A norm-conserving pseudo-potential for C (2s$^{2}$ 2p$^{2}$)\cite{lin93},
planewaves up to a cutoff of $800$ eV,
and $m \times m \times 1$ Monkhorst-Pack\cite{monk} meshes 
($m= 3, 1$ for $n=4, 6,$ respectively) make the other
important ingredients of our calculations. 
Spin-polarized calculations have been performed to take into
account systems with an odd number of electrons.
By computing selected configurations with greater accuracy, 
we estimate computational errors on total energies as $\pm 0.02$ eV, 
while energy differences are given with $\pm 0.01$ eV.
Accuracy of binding energies, however, depend on the different approximations
in the model, notably the choice of the exchange and correlation functional, 
leading to discrepancies in absolute values
($\sim 1$ eV for LDA and $\sim 0.5$ eV for PBE). 
This offset does not affect significantly the behaviour of the
chemisorption energy vs the external stress, therefore not interfering
in the main conclusions of this work.
Geometrical configurations where considered converged when the maximum 
remnant displacement of atoms was less than 0.001 {\AA}, and the maximum 
residual force on any atom was less than 0.01 eV/{\AA}. 
Under these conditions, the single graphene layer shows an optimum configuration
for a honeycomb lattice with carbon-carbon distances of $1.405$ {\AA}
and $120^{0}$ angles where residual forces are less than $10^{-6}$ eV/{\AA} 
and residual stresses are below $0.003$ GPa.
Clusters calculations using a MIDI basis and B3LYP\cite{gamess} yield
a similar C-C distance of $1.420$ {\AA}. 

To test the chemical activity of graphene we consider adsorption of
atomic hydrogen. Choosing a simple probe to study the chemical functionalization
of graphene layers has obvious advantages and has been shown to be 
useful to study defects on these layers\cite{boukhvalov08}. 
A MIDI/B3LYP model chemistry for H adsorbed on top the 
central C on a finite cluster (C$_{73}$H$_{22}$
yields a binding energy of $-0.19$ eV. 
The same calculation with LDA yields a binding energy of
$-0.95$ eV, which reflects the very well known tendency of 
local methods to overestimate binding energies for a large set of molecules (G1)
including many with similar C-C and C-H bonds\cite{b3lyp}. Use of 
the Perdew-Burke-Ernzerhof functional (PBE\cite{pbe}) 
improves the value to $-0.66$ eV but still is too large.
Results on an extended periodic system ($4 \times 4$) using a planewaves basis
are remarkably similar to the ones derived from finite clusters 
($-1.06$ eV and $-0.70$ for LDA and PBE, respectively).
As it has been extensively argued in the literature, this problem is not likely
to be solved by a gradient corrected approximation\cite{b3lyp}, neither
the revised-PBE\cite{rpbe} ($-0.63$ eV) nor Perdew-Wang\cite{pw91} 
($-0.67$ eV) get much closer to a realistic value.
Other authors working on similar approaches have already reported
similar too large binding energies for H on graphene\cite{boukhvalov08prb}.
The small binding energy of H on graphene obtained with a more accurate
hybrid functional can be understood from
the balance between the gain associated with the formation
of a C-H covalent bond and the loss of $\pi$ bonding around the
involved C atom.

As commented above, in this work we are more concerned with the variation of 
the chemisorption energy with stress than with its absolute value.
This is shown in Fig~\ref{Fig2} where to make easier
the comparison of slopes the LDA values obtained from a periodic
model has been corrected by a constant offset ($0.866$ eV).
A significant increase in the binding energy with the C-C stretch 
(almost linear) is seen in the range between typical C-C bonds in 
graphene ($0$\%) and a typical C-C single bond ($10$\%). 
It is interesting to notice that quite different models predict
a similar variation for the binding energy making the result robust
to the approximations involved.
Predicted geometrical parameters are quite insensitive to the 
particulars of the model too (Table~\ref{TabHGRA}).
The main difference across these models is the buckling of C$_{1}$. 
This buckling is largely related to the elastic energy stored in the
substrate by its quasi-pyramidal deformation and it shows a long-range
dependence that makes it sensitive to the specific boundary conditions
(e.g., compare the adsorption of single H on $4 \times 4$ with the 
adsorption of two H on different sides of a $6 \times 6$, the later
admitting more easily a lattice distortion because (i) H are farther
apart, and (ii) being located on opposite sides of the layer the
concave and convex distortions of the lattice meet better at the 
central symmetrical node line).
These distortions can seed the nucleation of topological disorder at
long distances, as can be seen in our larger cluster, where
the bond lengths relax to its equilibrium value
in an oscillatory way, reaching the boundary of the cluster.

Atomic H interacts weakly with a single graphene layer due to the robust
sp$^{2}$+$\pi$ bonds holding the layer. 
Standard DFT calculations using a local (LDA, PBE) functional for
exchange and correlation overestimate the binding energy by a factor
$\sim 5$ to $3$ over values obtained with a hybrid functional (B3LYP).
The energy depends linearly on the external stress and the slope
is well reproduced independently of the choosen exchange and 
correlation functional.
Tensile external stresses weaken
the extended $\pi$ orbital bonding activating an incipient dangling
bond that can bind strongly to the H atom. 
Under tensile stress of $\sim 20$ N m$^{-1}$ (half-way the breaking limit of 
the layer, equivalent to a C-C stretch of $\sim 10$ \%),
the graphene layer becomes $\sim 5$ times more reactive.
This is a reversible effect that can be switched on and off by modulating
the external stress.

This work has been financed by the Spanish
CICYT (MAT-2005-3866 and MAT-2006-03741), 
and MEC (CONSOLIDER NANOSELECT and NAN2004-09813-C10-08). 
We acknowledge the use of the Spanish Supercomputing
facilities.

\newpage


\newpage

\begin{figure}
\caption{(color online)
H (red) adsorbed on a cluster of atoms (C$_{73}$H$_{22}$)
forming a honeycomb lattice 
(carbon dangling bonds on the border have been saturated with hydrogen atoms).
The adsorption of H disturbs the planarity of the cluster and changes
the C-C distances around the adsorption site over a large distance.
}
\label{FigHC73H22}
\end{figure}

\begin{figure}
\caption{(color online)
Two H atoms (red) adsorbed on both sides of a $6 \times 6$ cell 
representing a 2D infinite system where the chemisorption problem
is solved using a plane-waves extended basis set.
Notice the buckling induced by the adsorption of H affecting
C-C bond lengths located at distances comparable to the size
of the cell.
}
\label{Fig6x6}
\end{figure}

\begin{figure}
\caption{(color online)
Binding energy (eV) of H on graphene vs. C-C stretch (\%) calculated for 
(a) cluster in Fig.~\ref{FigHC73H22}), and MIDI/B3LYP chemistry (circles 
and solid line), and
(b) a periodic $4 \times 4$ unit cell using planewaves and LDA 
(triangles and dashed line). 
Lines are least-square fits to guide the eye.
The LDA result has been corrected by an offset, $0.866$ eV, 
to allow the comparison of slopes.
}
\label{Fig2}
\end{figure}

\begin{table}
\caption{
Comparison of geometrical parameters for the three models 
considered for H adsorbed on graphene. 
In the $6 \times 6$ two H have been adsorbed on both sides of the layer
separated by $11.29$ {\AA}. 
The following parameters are listed: 
buckling of the C atom binding directly to the adsorbed H ($\Delta z$ C$_{1}$), 
the length of this bond (H-C$_{1}$), 
the angle defined by H, C$_{1}$ and C$_{NN}$ ($\alpha$), 
and the distance from C$_{1}$ to its nearest-neighbours (C$_{NN}$).
}
\begin{tabular}{lcccc}
\hline
MODEL & $\Delta z$ C$_{1}$ ({\AA}) & H-C$_{1}$ ({\AA}) & $\alpha$ (deg)  & C$_{1}$-C$_{NN}$ ({\AA}) \\
\hline
 C$_{73}$H$_{22}$ & 0.31 & 1.13 & 102$^{o}$ & 1.50  \\
 $4 \times 4$     & 0.41 & 1.13 & 103$^{o}$ & 1.48  \\
 $6 \times 6$     & 0.54 & 1.13 & 103$^{o}$ & 1.48  \\
\hline
\end{tabular}
\label{TabHGRA}
\end{table}

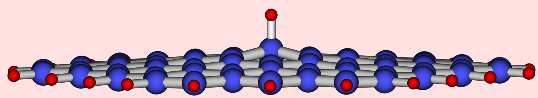
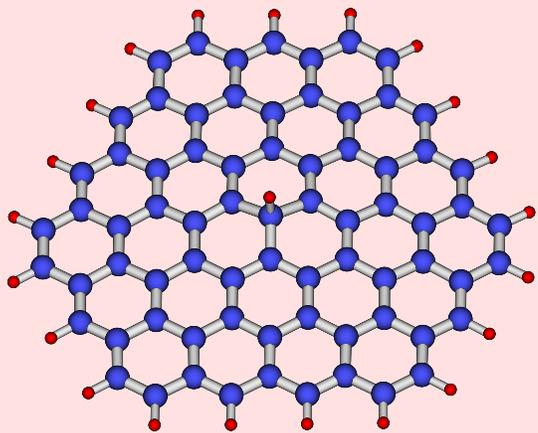
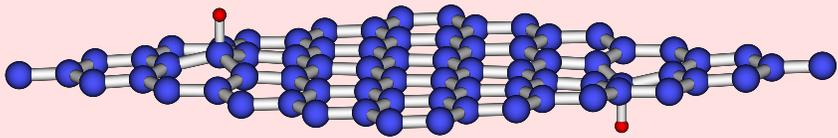
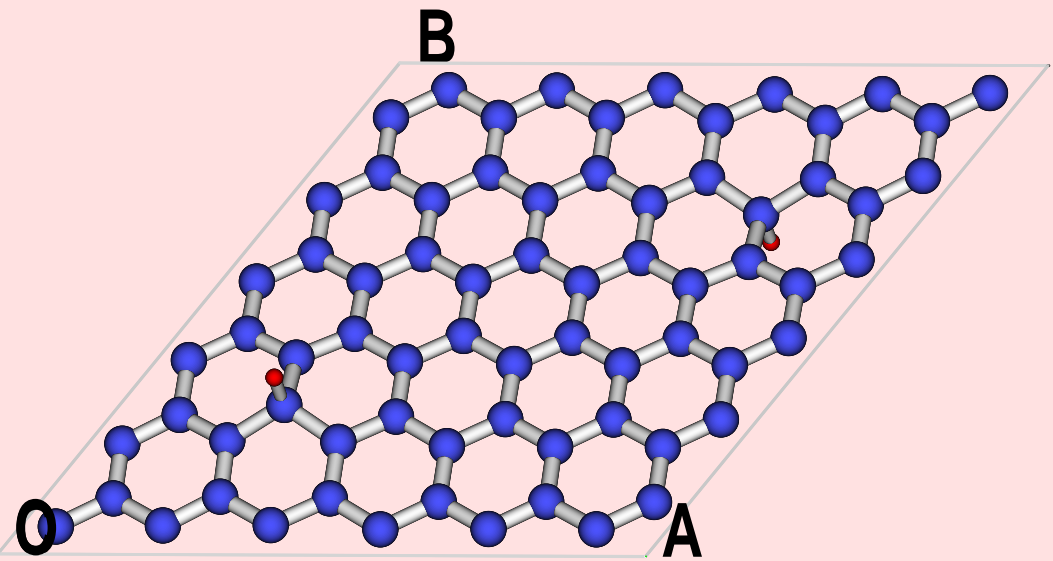
\end{document}